# A Wearable Social Interaction Aid for Children with Autism

*Authors: Nick Haber, Catalin Voss, Jena Daniels, Peter Washington, Azar Fazel, Aaron Kline, Titas De, Terry Winograd, Carl Feinstein, Dennis P. Wall*

With most recent estimates giving an incidence rate of 1 in 68 children in the United States [1], the autism spectrum disorder (ASD) is a growing public health crisis.

Many of these children struggle to make eye contact, recognize facial expressions, and engage in social interactions [2]. Today's standard for treatment of the core autism-related deficits focuses on a form of behavior training known as Applied Behavioral Analysis. To address perceived deficits in expression recognition, ABA approaches routinely involve the use of prompts such as flash cards for repetitive emotion recognition training via memorization [3]. These techniques must be administered by trained practitioners and often at clinical centers that are far outnumbered by and out of reach from the many children and families in need of attention. Waitlists for access are up to 18 months long, and this wait may lead to children regressing down a path of isolation that worsens their long-term prognosis [5]. There is an urgent need to innovate new methods of care delivery that can appropriately empower caregivers of children at risk or with a diagnosis of autism, and that capitalize on mobile tools and wearable devices for use outside of clinical settings.

The Autism Glass Project at Stanford University represents an attempt at such an innovation, using technology that need not be guided by a trained practitioner in order to create a learning experience within the natural contexts of expression recognition. We have developed an artificial intelligence tool [7,8,9,10] for automatic facial expression recognition that runs on smart glasses and delivers instantaneous social cues to people with autism. The computer vision system leverages the glasses' outward facing camera to read a person's facial expressions by passing video data to an Android native app for immediate machine learning-based emotion classification. The system then gives the child wearer real-time social cues and records social responses, allowing, for instance, the measurement of incidence of face-to-face (if not eye) contact. Through a dedicated app, caregivers can then review and discuss auto-curated videos of social interaction recorded throughout the day. The app highlights moments in which expressive events occurred, allowing, for instance, a mother to navigate to and review with a child an incident in which she had gotten angry earlier in the day. This brings therapy out of the clinician's office and into the homes of families, enabling caregivers to deliver therapy that generalizes learned skills into everyday life in a scalable fashion.

The device serves dually as a therapeutic tool and as a system of data capture, giving social interaction data that we hope will allow us to better understand the disorder. Imagine diagnosis, or at least triage, that uses an automated analysis of a short video recording of the child's interactions. Quantifying social interaction (e.g. "What percentage of the time that the mother is speaking is the child looking in her direction?") will hopefully move us closer to such a standard. This continuous quantification of social interaction also opens the door for us to be able to automatically, continuously track patients' progress. Through, for example, head pose habits or the ability to recognize facial expressions reliably in games, care providers will be able to keep tabs on patients without their visit to a care facility. This will allow us to see if the intervention is working quickly – actually quantifying what today's qualitative questionnaire diagnostic tools are getting at – and adjust treatment if necessary.

We intend the device to be a learning tool, not a prosthesis – patients are not expected to need to wear the device indefinitely, but rather to gain skills in recognizing expressions without any help. This relates to a number of difficult human-computer interaction questions. How do we give feedback to users in a way that allows them to naturally associate the face with the expression, rather than rely on the device? How should the system give cues in a way that is minimally obtrusive both parties in the conversation? How can we make conversation in the presence of this device natural?

In-lab pilot data [6] from 43 children, 23 with autism and 20 typically-developing, shows that children will wear such a device and can use such feedback in order to interpret facial expression stimuli. Following this in-lab exploratory work, we launched a design study with 28 families impacted by Autism, allowing them to take the devices home and use them on a daily basis over a period of weeks. This enabled us to refine design choices and develop a behavioral program that puts therapy in the hands of caregivers, either within the family unit or as an augmentation to existing ABA therapy. We tracked progress with gold-standard outcome measures and the continuously gathered device data. Along with observations of parents and teachers – often blind to treatment – this data suggests that the therapy delivered by our system has potential for dramatic improvement that is more generalized to daily life on a quicker scale than ABA therapy. The study produced over 8,000 minutes of social video and sensor data that we hope will shed further light on Autism as a disorder itself through a fine-grained analysis of the diverse behavioral patterns of autism.

References:
[1] Christensen, D.L., et al., *Prevalence and Characteristics of Autism Spectrum Disorder Among 4-Year-Old Children in the Autism and Developmental Disabilities Monitoring Network*. J Dev Behav Pediatr, 2016. 37(1): p. 1-8.
[2] Landa, R.J., K.C. Holman, and E. Garrett-Mayer, *Social and communication development in toddlers with early and later diagnosis of autism spectrum disorders*. Arch Gen Psychiatry, 2007. 64(7): p. 853-64.
[3] Lovaas, O.I., *Teaching Individuals With Developmental Delays: Basic Intervention Techniques*. 2003: ERIC.

[4]Schreibman, L., et al., *Naturalistic Developmental Behavioral Interventions: Empirically Validated Treatments for Autism Spectrum Disorder*. J Autism Dev Disord, 2015. 45(8): p. 2411-28.
[5] Dawson, G., *Early behavioral intervention, brain plasticity, and the prevention of autism spectrum disorder*. Development and psychopathology, 2008. 20(03): p. 775-803.
[6] Peter Washington, Catalin Voss, Nick Haber, Serena Tanaka, Jena Daniels, Carl Feinstein, Terry Winograd, Dennis Wall. *A wearable social interaction aid for children with autism*. ACM annual conference on Human Factors in Computing Systems (CHI) Late-Breaking Work, 2016.
[7] Catalin Voss, Peter Washington, Nick Haber, Aaron Kline, Jena Daniels, Azar Fazel, Titas De, Beth McCarthy, Carl Feinstein, Terry Winograd, Dennis Wall. *Superpower glass: delivering unobtrusive real-time social cues in wearable systems.* Proceedings of the 2016 ACM International Joint Conference on Pervasive and Ubiquitous Computing: Adjunct, 2016.
[8] Peter Washington, Catalin Voss, Aaron Kline, Nick Haber, Jena Daniels, Azar Fazel, Titas De, Carl Feinstein, Terry Winograd, Dennis Wall. *SuperpowerGlass: A Wearable Aid for the At-Home Therapy of Children with Autism*. Proceedings of the ACM on Interactive, Mobile, Wearable and Ubiquitous Technologies, 2017.
[9] Jena Daniels, Jessey Schwartz, Nick Haber, Catalin Voss, Aaron Kline, Azar Fazel, Peter Washington, Titas De, Carl Feinstein, Terry Winograd, Dennis Wall. *5.13 Design and efficacy of a wearable device for social affective learning in children with autism*. Journal of the American Academy of Child & Adolescent Psychiatry, 2017.
[10] Catalin Voss, Nick Haber, Peter Washington, Aaron Kline, Beth McCarthy, Jena Daniels, Azar Fazel, Titas De, Carl Feinstein, Terry Winograd, Dennis Wall. *Designing a Holistic At-Home Learning Aid for Autism.* CHI Autism Technssology Workshop, 2016.

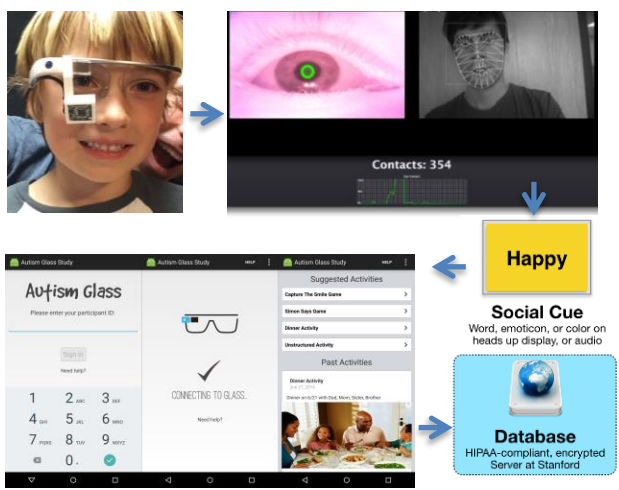

**Autism Glass system for at home therapy** The prototype consists of Google Glasses paired with a phone, with app displayed above. The device tracks faces and recognizes facial expressions, delivering social cues to the wearer. Video data is recorded and securely centralized.